\documentclass[%
reprint,
superscriptaddress,
noeprint,
amsmath,
amssymb,
pra,
]{revtex4-2}

\usepackage{xcolor}
\usepackage{graphicx}
\usepackage{dcolumn}
\usepackage{bm}
\usepackage{hyperref}
\hypersetup{colorlinks, allcolors=blue}
\usepackage[version=4]{mhchem}
\usepackage{siunitx}

\newcommand{\mb}[1]{{\textcolor{black} {#1}}}

\DeclareSIUnit\angstrom{\protect \text {Å}}

\begin{document}


\title{\mb{Direct observation of 2DEG states in shallow Si:Sb $\delta$-layers}}%

\author{Frode S. Strand}
\affiliation{Center for Quantum Spintronics, Department of Physics, Norwegian University of Science and Technology (NTNU), NO-7491 Trondheim, Norway.}

\author{Simon P. Cooil}
\affiliation{Department of Physics and Centre for Materials Science and Nanotechnology, University of Oslo (UiO), Oslo, 0318, Norway}

\author{Quinn T. Campbell}
\affiliation{Center for Computing Research, Sandia National Laboratories, Albuquerque, New Mexico 87185, USA}

\author{John J. Flounders}
\affiliation{Department of Physics and Astronomy, Interdisciplinary Nanoscience Center, Aarhus University, 8000, Aarhus C, Denmark}

\author{Håkon I. Røst}
\affiliation{Department of Physics and Technology, University of Bergen, 5007 Bergen, Norway}

\author{Anna Cecilie Åsland}
\affiliation{Center for Quantum Spintronics, Department of Physics, Norwegian University of Science and Technology (NTNU), NO-7491 Trondheim, Norway.}

\author{Alv Johan Skarpeid}
\affiliation{Department of Physics and Centre for Materials Science and Nanotechnology, University of Oslo (UiO), Oslo, 0318, Norway}

\author{Marte P. Stalsberg}
\affiliation{Department of Physics and Centre for Materials Science and Nanotechnology, University of Oslo (UiO), Oslo, 0318, Norway}

\author{Jinbang Hu}
\affiliation{Department of Physics, Norwegian University of Science and Technology (NTNU), NO-7491 Trondheim, Norway}

\author{Johannes Bakkelund}
\affiliation{Department of Physics, Norwegian University of Science and Technology (NTNU), NO-7491 Trondheim, Norway}

\author{Victoria Bjelland}
\affiliation{Department of Physics, Norwegian University of Science and Technology (NTNU), NO-7491 Trondheim, Norway}

\author{Alexei B. Preobrajenski}
\affiliation{MAX IV Laboratory, Lund University, 22100 Lund, Sweden}

\author{Zheshen Li}
\affiliation{Department of Physics and Astronomy, Aarhus University, Aarhus C, 8000, Denmark}

\author{Marco Bianchi}
\affiliation{Department of Physics and Astronomy, Interdisciplinary Nanoscience Center, Aarhus University, 8000, Aarhus C, Denmark}

\affiliation{Elettra - Sincrotrone Trieste, AREA Science Park, 34149, Trieste, Italy}


\author{Jill A. Miwa}
\affiliation{Department of Physics and Astronomy, Interdisciplinary Nanoscience Center, Aarhus University, 8000, Aarhus C, Denmark}

\author{Justin W. Wells}
\email[Corresponding author: ]{j.w.wells@fys.uio.no}
\affiliation{Center for Quantum Spintronics, Department of Physics, Norwegian University of Science and Technology (NTNU), NO-7491 Trondheim, Norway.}
\affiliation{Department of Physics and Centre for Materials Science and Nanotechnology, University of Oslo (UiO), Oslo, 0318, Norway}

\date{\today}

\begin{abstract}
We investigate the electronic structure of high-density layers of Sb dopants in a silicon host, so-called Si:Sb $\delta$-layers. We show that, in spite of the known challenges in producing highly confined Sb $\delta$-layers, sufficient confinement is created such that the lowest conduction band states ($\Gamma$ states, studied in depth in other silicon $\delta$-layers), become occupied and can be observed using angle-resolved photoemission spectroscopy. 
\mb{The electronic structure of the Si:Sb $\delta$-layers closely resembles that of Si:P systems, where the observed conduction band is near-parabolic and slightly anisotropic in the $\mathbf{k}_\parallel$ plane.
The observed $\Gamma$ state extends $\sim \SI{1}{\nm}$ in the out-of-plane direction, which is slightly wider than the 1/3 monolayer thick dopant distribution. %
}
\mb{This is caused by a small segregation of the dopant layer, which is nevertheless minimal when comparing with earlier published attempts.}
\mb{Our results serve to demonstrate that Sb is still a feasible dopant alternative for use in the semiconductor $\delta$-layer platform, providing similar electronic functionality to Si:P systems. Additionally, it has the advantages of being less expensive, more controllable, safer to handle, and more compatible with industrial patterning techniques. Si:Sb is therefore a viable platform for emerging quantum device applications. }
\end{abstract}

\maketitle

\section{Introduction}\label{sec:intro}

Over the past few decades, the process of silicon $\delta$-doping has seen a lot of activity due to its potential as a platform for quantum computing architectures \cite{Fuechsle2012transistor, Veldhorst2015twoqubit, Zwanenburg2013silicon}. 
To this end, considerable effort has been devoted to understand the electronic properties of the resulting two-dimentional electron gases (2DEGs) and develop the fabrication processes to improve the confinement and stability of the dopant layers \cite{Polley2013exploring}.
In recent years, by far the most extensively studied platform has been Si:P $\delta$-layers, in which phosphorus is used to create a sharp and dense \emph{n}-type dopant profile in a silicon host. 
Many details about the electronic structure and physical properties have been revealed both theoretically \cite{Carter2009electronic, Lee2011electronic, Carter2011phosphorus} and experimentally \cite{Miwa2013direct, Miwa2014valley, Mazzola2014disentangling, Holt2020observation, Mazzola2020thesub, Constantinou2023momentum}. The popularity of phosphorus as a dopant is due to the self-limiting -- and thus, consistent achievable dopant density of the Si:P fabrication process, with relatively small dopant segregation \cite{Nutzel1995comparison}.
Other dopants have also been investigated, including arsenic \cite{Stock2020atomic} and bismuth \cite{Lundgren2023bismuth}. However, the fabrication of $\delta$-layers using these dopants all rely on toxic precursor gases, phosphine (\ce{PH3}), arsine (\ce{AsH3}), and bismuth trichloride (\ce{BiCl3}). Herein, we investigate elemental antimony (Sb) as a dopant specimen and show that a dense and sharp dopant profile can be created without the need for toxic precursors. \mb{Instead of gas phase deposition, Sb can be deposited on the substrate by simple sublimation from a solid Sb source.}

In fact, $\delta$-doping in silicon appears to have been first demonstrated using Sb as a dopant \cite{Zeindl1987growth}, and different variants of the Si:Sb system were studied by several groups well into the 1990s \cite{Li1989electrical, Ni1992delta, Gossmann1993dopant, Kruger1996characterization, Citrin1999geometric}. However, due to challenges with dopant diffusion, the collective attention of the community shifted by a large degree to the more promising Si:P system mentioned above.
\mb{Nevertheless, building on the knowledge gained on Si:P and recent advances in experimental characterisation techniques, we revisit the Si:Sb system and show that it is indeed possible to create extremely dense and narrow dopant profiles (i.e. metallic over $\sim \SI{1}{\nm}$), comparable to Si:P $\delta$-layers. These profiles give rise to sufficient confinement for the conduction band minimum to become occupied. We perform density functional therory (DFT) calculations and angle-resolved photoemission spectroscopy (ARPES) measurements and show, for the first time, the occupied band structure of the buried Si:Sb system.} 

\begin{figure*}
    \centering
    \includegraphics[width=\textwidth]{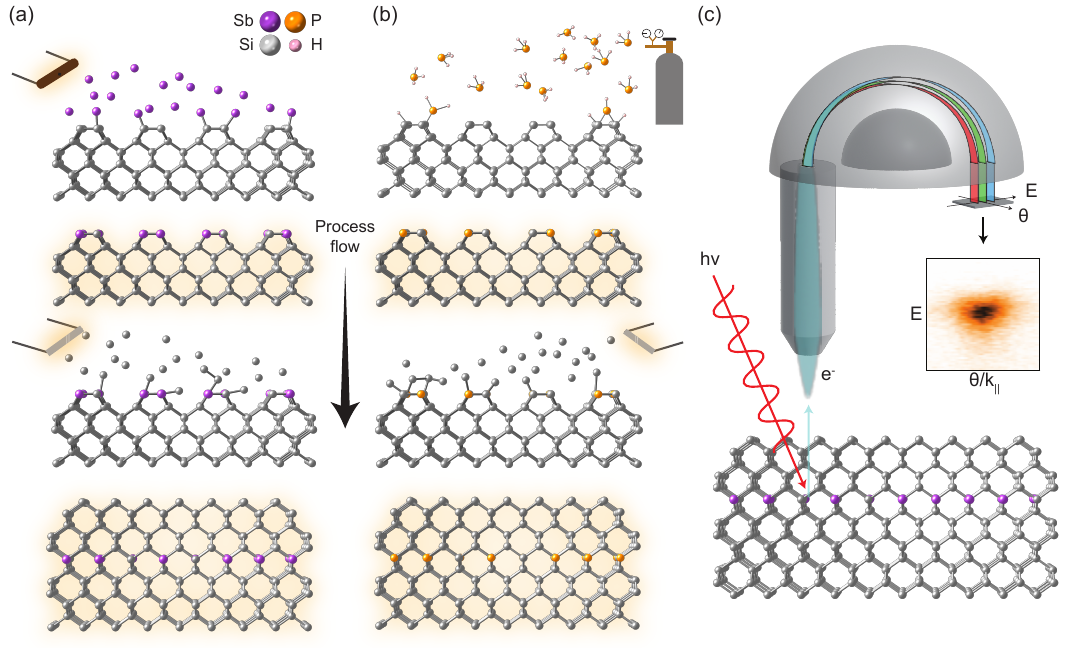}
    \caption{Schematic of the $\delta$-layer fabrication process and measurement. (a) Fabrication using Sb as a dopant. Solid Sb is thermally evaporated from a source onto a $(2 \times 1)$-reconstructed Si(001) surface (top). Subsequent annealing to $\sim \SI{350}{\celsius}$ for a few seconds incorporates the dopants into the surface (middle), and is then encapsulated by a layer of Si with an extra annealing step to $\sim \SI{450}{\celsius}$ to recrystallise the surface (bottom). (b) The corresponding process for making the widely studied Si:P $\delta$-layer. \ce{PH3} is leaked into the vacuum chamber and is adsorbed onto the surface. As before, annealing to $\sim \SI{350}{\celsius}$ results in P incorporation into the surface, and the following steps are nearly identical. (c) Illustration of the XPS and ARPES experiments. Electrons in the sample are photoexcited and then measured by a hemispherical analyser, providing information about the chemical environment of the sample as well as the electronic band structure. 
    }
    \label{fig:schematic}
\end{figure*}

A schematic comparison of Si:P and Si:Sb $\delta$-layer fabrication is shown in Fig.\ \ref{fig:schematic} to illustrate their practical differences. 
In Fig.\ \ref{fig:schematic}(a), Sb atoms are deposited from an evaporation source onto a clean Si surface held at room temperature (here shown as the $(2 \times 1)$-reconstructed (001) surface) in an ultra-high vacuum (UHV) environment. The sample is subsequently annealed to a moderate temperature (typically in the range 300 - \SI{400}{\celsius}) to incorporate the dopants into the surface \cite{Garni1999stm}. In the case of P doping, shown in Fig. \ref{fig:schematic}(b), 
the Si(001) surface is exposed to 1.125 Langmuirs (L) of \ce{PH3} gas (\SI{5e-9}{\milli\bar} for 5 minutes).
A similar annealing step is then performed, which results in the release of H and a $\sim 1/4$ monolayer (ML) coverage of P dopants \cite{Wilson2006thermal}. If a higher dopant density is desired, it is necessary to deposit more Si on top of the first layer to allow the \ce{PH3} dissociation and incorporation process to take place, or anneal to remove H such that the surface is reactivated \cite{McKibbin2014low,Mazzola2020thesub}. 
\mb{In contrast, Sb can be incorporated into the Si(001) surface in a two-dimensional manner up to about 0.5 ML, whereafter Sb growth likely changes to a Stranski-Krastanov growth mode \cite{Metzger1984antimony, Jewell2018low}.}

\mb{After the Sb or P dopants are incorporated into the surface, a Si encapsulation layer is grown on top and subsequently annealed to crystallise the epilayer. This final anneal temperature is more critical since overheating promotes segregation of the dopants \cite{Goh2004effect}}.

Herein, our grown Si:Sb $\delta$-layers are probed using ARPES, a surface-sensitive technique which gives access to the electronic band structure (Fig.\ \ref{fig:schematic}(c)). Photoemission from the buried $\delta$-layer is possible due to resonances that occur at certain photon energies (explained below) \mb{Still, the Si encapsulation still needs to be thin enough such that the probed electronic state has an appreciable amplitude near the surface, e.g. \SI{4}{\nm} \cite{Mazzola2018simultaneous}. }

\section{Methods}
$\left(2\times1\right)$-reconstructed Si(001) surfaces were prepared in vacuum by cycles of high-temperature flash-annealing up to $\sim \SI{1200}{\celsius}$. The surface cleanliness and structure were verified by X-ray photoelectron spectroscopy (XPS) and low-energy electron diffraction (LEED), respectively. \mb{Sample temperatures were monitored with a pyrometer with emissivity $\epsilon = 0.66$. Sb was deposited by thermal evaporation from a Ta crucible. The evaporation rate was controlled by varying the current through the crucible such that submonolayer thicknesses of Sb could be achieved in 20-30 minutes. The substrate was held at room temperature during Sb deposition.}
Next, the samples were annealed to $\SI{350}{\celsius}$ to promote Sb incorporation into the surface. Finally, a 0.5 - \SI{1.0}{\nm} thick Si encapsulation layer was deposited at room temperature and annealed to $\sim \SI{450}{\celsius}$ for a few seconds to crystallise the overlayer. 
XPS and ARPES measurements were carried out at the FlexPES beamline \cite{Preobrajenski2023} at MAX-IV Laboratory, Lund, Sweden, and the SGM3 beamline at the ASTRID2 synchrotron in Aarhus, Denmark \cite{Hoffmann2004, Bianchi2023}.

Electronic structure calculations were carried out using Quantum ESPRESSO \cite{Giannozzi2009quantum}. 
We used fully relativistic PAW pseudopotentials from the PSlibrary \cite{DalCorso2014pseudopotentials}
and the Perdew-Burke-Ernzerhof exchange-correlation functional \cite{Perdew1985density}. We used kinetic energy cutoffs of 50 Rydberg (Ry) and 200 Ry for the plane wave basis sets that define the Kohn-Sham orbitals and charge density, respectively. We used a 2$\times$2$\times$1 Monkhorst-Pack grid \cite{Monkhorst1976special} to sample the Brillouin zone in our initial self-consistent calculation and then a 4$\times$4$\times$1 Monkhorst-Pack grid for non self-consistent calculations before band structures were calculated. Unless otherwise noted, we allowed the atoms to undergo geometric relaxation at a fixed supercell shape until the interatomic forces fell below a threshold of \SI{50}{\milli\eV\per\angstrom}. The cell size was kept fixed to mimic embedding the $\delta$-doped layer in a larger bulk-like silicon crystal. We included spin-orbit coupling in all our calculations.

To calculate the electronic structure of $\delta$-layers, we created a 2$\times$2 Si(001) slab \SI{110}{\angstrom} thick, with a single layer of 1/4 ML concentration of Sb atoms at substitutional sites. The silicon lattice constant was set at \SI{5.468}{\angstrom}. Because we have used a plane wave DFT code that enforces periodic boundary conditions in the Kohn-Sham orbitals and potential, the modelled slab repeats in the (001) direction and the $\delta$-layer interacts with an infinite number of its images spaced with a period of \SI{110}{\angstrom}. Previous work has shown this distance to be sufficient to mitigate the effect of these spurious images on our results \cite{Campbell2023electronic}.

\section{Results and Discussion}

\begin{figure*}[!ht]
    \centering
    \includegraphics[width=\textwidth]{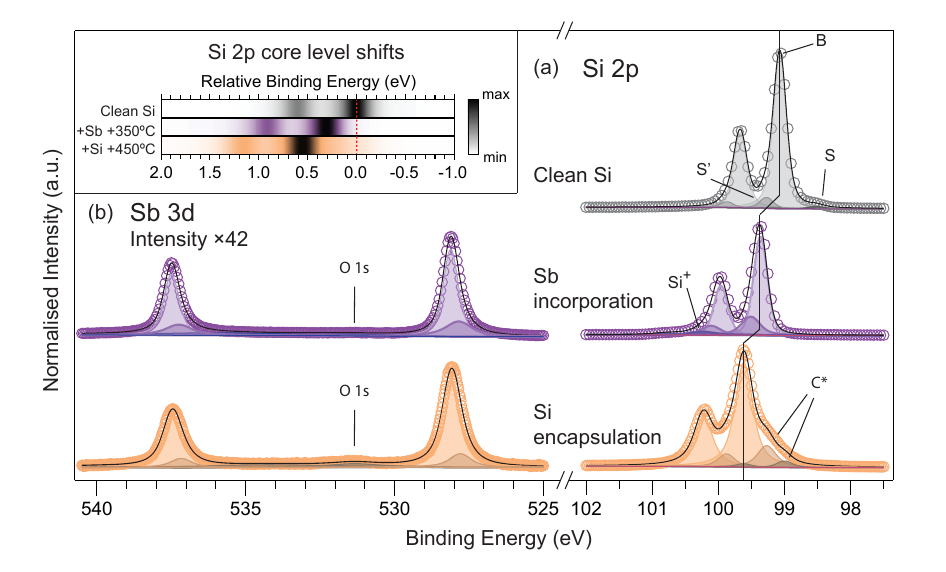}
    \caption{%
    XPS spectra for each preparation step, collected using \SI{590}{\eV} excitation energy and normal emission. (a) Si 2p core level spectra for clean Si (top, grey), with Sb incorporated into the surface (middle, purple) and the annealed $\delta$-layer (bottom, yellow). (b) Corresponding Sb 3d core level spectra. The Sb 3d intensities have been scaled up by a factor of 42 to facilitate comparison with the much more intense Si 2p spectra. Inset: Colourscale plots of the Si 2p core level highlighting the shift in binding energy relative to the primary component `B' for clean Si during sample preparation. The black part of the colourscale (illustrated for the top, grey spectrum) represents the peak in (a).  
    }
    \label{fig:xps}
\end{figure*}

In Fig.\ \ref{fig:xps}(a), the Si 2p core level from the clean Si(001) sample (top, grey) shows the familiar components consistent with a reconstructed ($2 \times 1$) surface \cite{Landemark1992corelevel}, with the most intense peak located at binding energy $E_\text{B} = \SI{99.04}{\eV}$, labelled `B' for `bulk'. 
The photon energy used (\SI{590}{\eV}) in the XPS measurements results in an inelastic mean free path (IMFP) on the order of \SI{1}{\nm}, thereby causing the signal to be dominated by subsurface core level electrons \cite{Hufner2003}. We therefore ascribe the dominant peak to bulk-like Si for all preparation steps, which is consistent with previous observations \cite{Lin1991dimer, Rost2023probing}.
The doublet at lower $E_\text{B}$ is associated with surface Si dimers (S) and the higher $E_\text{B}$ (S') is from second layer atoms \cite{Lin1991dimer, Landemark1992corelevel}. Quantification of the XPS spectra shows that a $\sim$ 1/3 ML of Sb was achieved upon deposition, similar to previous studies on Si:P $\delta$-layers with one or multiple rounds of \ce{PH3} dosing and P incorporation \cite{McKibbin2014low,Rost2023probing}. Assuming full dopant activation, this results in a nominal sheet carrier density $n \approx \SI{2.3d14}{\per\square\centi\metre}$. 
Subsequent annealing to \SI{350}{\celsius} shifts the whole Si 2p spectrum to larger $E_\text{B}$ by \SI{0.30}{\eV}. This is clarified by the colourscale plot in the inset in Fig.\ \ref{fig:xps}, showing the binding energy of the Si 2p core level relative to the `B' component for clean Si (top). Moreover, the annealing step diminishes the `S' component, indicating that the Si dimer bonds revert to a bulk-like configuration. In addition, a new component appears at $+\SI{0.86}{\eV}$ relative to `B'. This feature is ascribed to \ce{Si+} from small traces of oxygen contamination during Sb growth \cite{Hollinger1984probing}.
The corresponding O 1s component can be seen in the Sb 3d spectra in Fig.\ \ref{fig:xps}(b).

Upon deposition and subsequent annealing of a $\sim \SI{0.3}{\nm}$ thick Si encapsulation layer, the `B' component shifts further by +\SI{0.24}{\eV} (+\SI{0.54}{\eV} relative to clean Si in total). This indicates further dopant activation compared to the non-encapsulated Si-Sb surface. 
Two new components at lower $E_\text{B}$ are also formed, seen as a flat shoulder in the Si 2p core level. The origin of these components is not immediately clear, but a likely explanation is that the final annealing step causes some dopant segregation, creating a mixed, unreconstructed Si-Sb surface with a small degree of roughness \cite{Jiang1998strong}. 

\begin{figure*}
    \centering
    \includegraphics[width=0.95\textwidth]{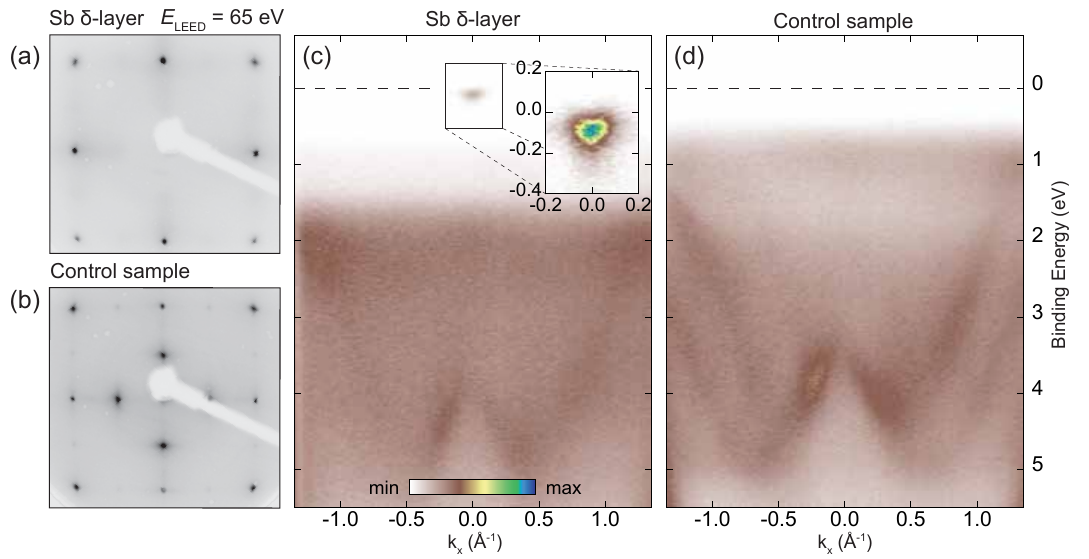}
    \caption{%
    LEED and ARPES data for a Si:Sb $\delta$-layer and a control sample. Panels (a) and (c) show surface crystal structure and electronic band dispersion in the K-$\Gamma$-K direction of a buried Sb $\delta$-layer, respectively. Panels (b) and (d) show identical datasets acquired following an identical preparation recipe to (a) whilst skipping the Sb deposition, i.e. Si overlayer growth with no $\delta$-layer formation. LEED and ARPES measurements were acquired using electron kinetic energy $E_\text{LEED} = \SI{65}{\eV}$ and photon energy $h\nu = \SI{107}{\eV}$.
    %
    }
    \label{fig:arpes}
\end{figure*}

A LEED pattern is shown in Fig.\ \ref{fig:arpes}(c), which shows that the final surface has a $1 \times 1$ structure with very faint $2 \times 1$ spots, consistent with this picture. It has been shown previously that an Sb-terminated Si(001) surface causes a relaxation in the Si substrate in which half of the third-layer atoms contribute to a lower $E_\text{B}$ component, given that there are Sb-Sb dimers on the surface \cite{DePadova1998identification}. We therefore label these C* in accordance with Ref. \onlinecite{DePadova1998identification}. The faint $2 \times 1$ spots could indicate that there is a degree of dimerisation on the surface. 

Upon annealing of the encapsulated $\delta$-layer structure, the Sb 3d core level intensity increases back to a similar level as to the non-encapsulated case. In other words, our $\delta$-layer is somewhat broadened, or ``smeared out'', in the out-of-plane direction. Broadening of the $\delta$-profile during growth is a well-known problem in $\delta$-layer fabrication, especially with Sb dopants \cite{Blacksberg2005}, so it is not surprising that this is observed here. In our case, however, the dopants seem to only moderately diffuse towards the surface because the observed shift in the `B' component of the Si 2p core level is indicative of strong band bending due to the presence of active dopants well beneath the surface. It has been shown before that the surface concentration of Sb depends on the thickness of the Si encapsulation layer. For encapsulation layer thicknesses below \SI{1}{\nm}, a non-zero concentration of dopants at the surface can be expected even at temperatures as low as \SI{130}{\celsius} \cite{Kimura1996anomalous}. Since we crystallise the Si overlayer using a rapid thermal annealing (RTA) method for only a few seconds at a moderate temperature ($< \SI{450}{\celsius}$), the dopant diffusion should be limited \cite{Gossmann1993doping}.

Despite the indications of Sb segregation, the buried layer still contains a sufficiently high density of dopants to occupy the conduction band minimum (CBM). \mb{An overview ARPES spectrum of an Sb-doped $\delta$-layer, measured with a photon energy of \SI{107}{\eV} along the K-$\Gamma$-K direction, is shown in Fig.\ \ref{fig:arpes}(c). A clear feature (highlighted in the inset) is apparent just below the Fermi level ($E_\text{F}$), consistent with $\delta$-states observed in previous reports on Si:P systems \cite{Miwa2013direct, Miwa2014valley, Mazzola2014determining, Mazzola2014disentangling, Mazzola2020thesub, Constantinou2023momentum}.}
The corresponding LEED pattern is shown in Fig.\ \ref{fig:arpes}(a), displaying a faintly streaked $1 \times 1$ pattern with very weak and diffuse $2 \times 1$ spots. A comparison to a control sample is made in \ref{fig:arpes}(b) and \ref{fig:arpes}(d): this sample was prepared by following the exact same steps as the $\delta$-doped sample represented in (c), but without depositing any Sb before adding the Si overlayer. It retains a $2 \times 1$ LEED pattern as with a flashed \ce{Si}(001) surface. The valence band maximum (VBM) of the $\delta$-doped sample is shifted towards higher binding energy by $\approx \SI{0.6}{\eV}$, which is consistent with the core-level shifts observed in Fig.\ \ref{fig:xps}.

\begin{figure*}
    \centering
    \includegraphics[width=\textwidth]{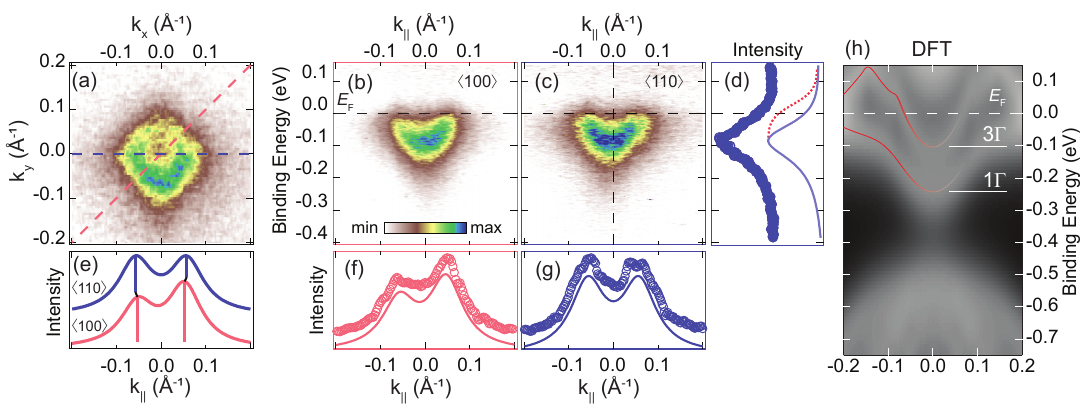}
    \caption{Fermi surface and selected ARPES measurements of a buried Sb $\delta$-layer. 
    (a) Constant energy surface showing the photoemission intensity for the 1$\Gamma$ state as a function of $\mathbf{k}_\text{x}$ and $\mathbf{k}_\text{y}$ at \SI{36}{\eV} photon energy. The pink and blue lines represent the $\left[100\right]$ and $\left[110\right]$ crystallographic directions and are the directions used to extract the $E$ vs. $\textbf{k}$ data shown in (b) and (c), respectively. (d) An EDC is extracted from (c) and fitted with a single Lorentzian modulated by a Fermi function (blue and red curves, respectively). (e) Lorentzian fits of the MDCs shown in (f) and (g), highlighting the difference in peak separations for $\left[ 100 \right]$ and $\left[ 110 \right]$ directions. This difference is $\approx \SI{0.01}{\per\angstrom}$. (f, g) MDCs extracted near the Fermi level from (b) and (c), respectively, fitted with Lorentzian doublets. (h) Simulated spectral function of a 1/4 ML Si:Sb $\delta$-layer with bare $1\Gamma$ and $3\Gamma$ bands overlaid in red.
     }
    \label{fig:detail}
\end{figure*}

Detailed ARPES spectra of the $\delta$-state is shown in Fig.\ \ref{fig:detail}, where (a) shows a constant energy surface map measured at $E_\text{F}$ with pink and blue dashed lines to indicate the $\left[ 100 \right]$ and $\left[ 110 \right]$ directions shown in (b) and (c), respectively. Corresponding momentum distribution curves (MDCs) have been extracted (integration width \SI{15}{\milli\eV} at $E_\text{F}$) as indicated by the dashed lines in each panel and shown in \ref{fig:detail}(f) and (g), where they are fitted with two Lorentzian lineshapes. The fits of the two MDCs are plotted together in \ref{fig:detail}(e). Clearly, the $\delta$-state is not quite isotropic in the in-plane wavevector $\textbf{k}$ as indicated by the diamond-shaped Fermi contour in (a). This is further emphasised by the difference in the separation of the two Lorentzian components in each case. The asymmetry in the MDC intensity in Fig.\ \ref{fig:detail}(f) can be attributed to the geometry of the photoemission experiment: 
the emission angle is measured to acquire $E$ vs. $\textbf{k}_\text{x}$ and scanned through $\textbf{k}_\text{y}$ by rotating the sample relative to the detector (see Fig.\ \ref{fig:schematic}), thus changing the photoionisation matrix elements of the photoemission process \cite{Damascelli2004arpes}.
It should be noted that the fit did not improve when including more than one doublet in the fit function for the MDC.
An energy distribution curve (EDC) extracted along $\textbf{k}_\parallel = 0$ (vertical dashed line in Fig.\ \ref{fig:detail}(c)) is shown in \ref{fig:detail}(d) and fitted with a single Lorentzian component (blue line) modulated by a Fermi-Dirac function (grey line). The EDC is has its intensity maximum about \SI{80}{\milli\eV} below $E_\text{F}$.

A simulated ARPES dataset for an Sb $\delta$-layer is shown in Fig.\ \ref{fig:detail}(h), assuming a sharp and ordered 1/4 ML profile.
\mb{Here, the spectral function for the bare bands have been simulated, and realistic $E$ and $\textbf{k}$ resolutions have been applied to facilitate direct comparison to our experimental data. It should be kept in mind that band gaps tend to be underestimated in semiconductors with our DFT approach \cite{Perdew1985density}. Nevertheless, the model serves to qualitatively illustrate the expected impact of the dopant layer on the conduction band states.}
\mb{As can be seen in the figure, this model produces a band structure in which two states (red curves) separated by $\sim \SI{140}{\milli\eV}$ appear in the conduction band. 
These bands are known as the $\Gamma$ states. In comparison, three distinct bands have been observed to be occupied in Si:P (labelled $1\Gamma$ through $3\Gamma$) \cite{Mazzola2020thesub}. In the Si:P system, the higher $E_\text{B}$ band is seen to comprise two states with a very small separation ($1\Gamma$-$2\Gamma$) which, in turn, is well separated from the third state ($3\Gamma$). }

\mb{Note that our present DFT model does not account for the origin of the $1\Gamma$-$2\Gamma$ splitting, but the discussion remains largely the same as the general features are similar.
The two $\Gamma$ states in \ref{fig:detail}(h) both appear below $E_\text{F}$, which contrasts with the ARPES measurements (Fig.\ \ref{fig:detail}(b)-(d)) where only one state is resolved. This suggests that only the $1\Gamma$-$2\Gamma$ state is occupied.} 
The discrepancy between the measured and calculated CBM locations is likely due to the fact that the calculation assumes a perfectly confined $\delta$-layer. The real $\delta$-layer is more diffuse, thus shifting the Fermi level further up due to a higher dopant concentration in the vicinity of the layer.

\mb{The fact that we only see one state in the ARPES spectra is not surprising, since the expected separation between the $1\Gamma$ and $3\Gamma$ states is relatively large. From our experimentally observed level of doping, this would put the $3\Gamma$ state above $E_\text{F}$.} This large separation occurs because the confinement potential is very sharp, which is only achieved for an extremely sharp dopant profile. As was discussed previously, the actual Sb $\delta$-layer is most likely broadened to some degree, which should result in a less sharp confinement potential. This phenomenon has been shown experimentally with Si:P $\delta$-layers by Holt \emph{et al.} \cite{Holt2020observation}, where the separation of the bands were measured as a function of dopant layer thickness. In the single layer case (i.e. 1/4 ML), their measured separation between the CBM (i.e. the $1\Gamma$-$2\Gamma$ states) and $3\Gamma$ was at least \SI{230}{\milli\eV}. The bands remained well separated until they began to overlap as the dopant layer thickness was increased to \SI{4}{\nano\metre}. 

\mb{A similar trend can be expected for the Si:Sb system. Our observed CBM is about \SI{80}{\milli\eV} below $E_\text{F}$, which gives a lower bound for the energy separation to the next band ($3\Gamma$). Using the calculated separation of \SI{140}{\milli\eV} for a sharp 1/4 ML Sb layer as a starting point, the width of the actual $\delta$-layer profile should not be thicker than $\sim \SI{1}{\nm}$. Otherwise, the $3\Gamma$ state would become occupied due to the expected smaller energy separation. This then serves as a rough upper estimate of the perpendicular width of the confinement potential. }

Another factor affecting the number of $\Gamma$ states seen in our experiments is the symmetry of the dopant profile. The aforementioned smearing of the dopant profile is most likely asymmetric with a tail towards the surface due to some degree of dopant segregation. \mb{The effect of such asymmetry is to reduce the $1\Gamma$-$2\Gamma$ valley splitting and \emph{increase} the $1\Gamma$-$3\Gamma$ separation \cite{Mazzola2020thesub}, and could be a contributing factor as to why we do not see the $1\Gamma$-$2\Gamma$ splitting in our data. The same effect is expected due to the fact that the measured dopant concentration was slightly larger compared to the simulation.}

One would expect that spin-orbit coupling (SOC) effects could play a larger role in the bandstructure when using a heavier dopant element like \ce{Sb} instead of \ce{P}. The relativistic DFT calculation presented here indeed allows us to evaluate the impact of SOC on the band structure. The calculated SOC at the $E_\text{F}$ crossing is nevertheless very small ($<\SI{5}{\milli\eV}$), and well below the experimental detection limit. This manifests only as a very slight separation in the bare bands overlaid on the spectral image in Fig.\ \ref{fig:detail}(h). Although the SOC is small, it is an order of magnitude larger than in Si:P $\delta$-layers (previously reported to be $\approx \SI{0.8}{\milli\eV}$ \cite{Mazzola2020thesub}). Furthermore, the possibility of selecting the SOC strength (by choice of dopant species) indirectly allows some control of the spin relaxation time -- which is of central importance in spin-based quantum computation platforms \cite{Chen:2021}.

\begin{figure*}
    \centering
    \includegraphics[width=\textwidth]{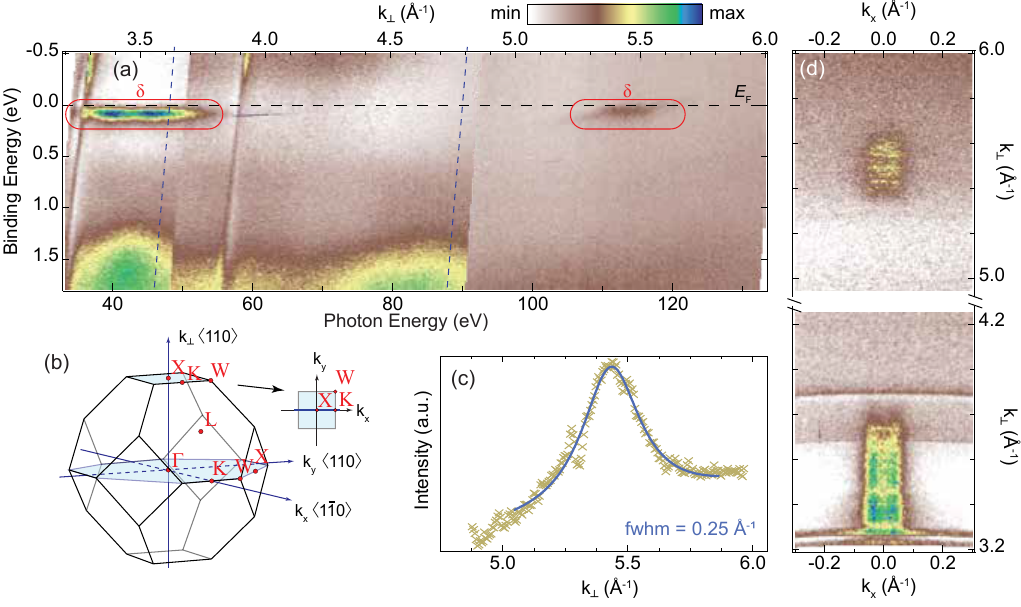}
    \caption{Photon energy dependent ARPES measurements. (a) Photoemission intensity as a function of $\textbf{k}_\perp$, corresponding to a photon energy range of 32 - \SI{132}{\eV}. The $\delta$-states appear near the Fermi level (dashed line) and are centred near 3.5 and \SI{5.5}{\per\angstrom}. (b) Silicon bulk Brillouin zone with definitions of axes $\textbf{k}_\text{x}$, $\textbf{k}_\text{x}$ and $\textbf{k}_\perp$ and some of the high symmetry points. The plane through the X point is shown with a blue line along $\textbf{k}_\text{x}$, indicating the sample alignment with the detector. (c) Integrated intensity of the $\delta$-layer state around \SI{5.5}{\per\angstrom}, fitted with an expression on the form $I \propto (\chi - 1)^2/(1 + \chi^2 - 2\chi\cos(\mathbf{k}_\perp a - \pi))$ (details in main text). (d) Constant binding energy slice at the Fermi level showing the $\delta$-layer states as in (a). }
    \label{fig:energyscan}
\end{figure*}

In order to probe the dispersion along the surface normal ($\textbf{k}_\text{x} = \textbf{k}_\text{y} = 0$), an ARPES spectrum was acquired whilst varying the photon energy, thus giving access to $\textbf{k}_\perp$. To convert from photon energy units to $\textbf{k}_\perp$, we assumed that the final states were free-electron-like, which allowed us to use the formula 
\begin{equation}
    \textbf{k}_\perp \approx \sqrt{\frac{2m}{\hbar^2} [V_0 + h\nu - E_\text{B} - \Phi]},   
\end{equation}
where $m$ is the electron rest mass, $V_0$ is the inner potential, $E_\text{B}$ the binding energy and $\Phi$ the surface work function of the sample \cite{Miwa2013direct}. 

The photoemission intensity as a function of $\textbf{k}_\perp$ is shown in Fig.\ \ref{fig:energyscan}(a). The $\delta$-state appears near $\textbf{k}_\perp = 3.5$ and \SI{5.5}{\per\angstrom}, and being metallic, determines the Fermi level (marked with a dashed line). These positions correspond to photon energies of roughly \num{36} and \SI{108}{\eV}. The clear lack of dispersion in $\textbf{k}_\perp$ confirms the 2D nature of the state, which is further demonstrated by the shape of the Fermi contours shown in Fig.\ \ref{fig:energyscan}(d). Here, each resonance of the $\delta$-state appears as two parallel rods or a square in the higher $\textbf{k}_\perp$ case (due to the lower signal-to-noise ratio) when the constant energy slice is taken at the Fermi level. 
Note that the intensities with apparent linear dispersions that appear above $E_\text{F}$ near $\textbf{k}_\perp = 3.3$ and \SI{3.9}{\per\angstrom} are simply excitations of the Sb 4d and Si 2p core levels due to second-order synchrotron light. 

It is possible to extract the real-space width of the initial electronic $\delta$-state by considering the variation in photoemission intensity as a function of $\textbf{k}_\perp$ \cite{Mazzola2014determining}. For photoemission to be possible from a 2D state, it has to occur via an electron final state which is well-matched in energy and momentum such that both quantities are conserved during photoexcitation and -emission. Moreover, the final state must be delocalised out-of-plane from the $\delta$-layer (i.e. a 3D state) to allow the excited electron to propagate to the surface. \mb{The Fourier transforms (FT) of the initial and final states in real space correspond to their $\textbf{k}_\perp$-distributions in reciprocal space, both of which have strongly peaked spectral weights \cite{Mazzola2014determining}.} By varying the photon energy in an ARPES experiment, the corresponding photoemission intensity will be resonantly enhanced where matching initial and final states overlap, resulting in a peaked intensity as a function of $\textbf{k}_\perp$ with a shape defined by their convolution. This is shown in Fig.\ \ref{fig:energyscan}(c), where the intensity of the $\delta$-state has been integrated and represented with a curve on the form $I \propto (\chi - 1)^2/(1 + \chi^2 - 2\chi\cos(\textbf{k}_\perp a - \pi))$. 
Here, $a$ is the Si lattice constant and $\chi$ is a parameter related to the bandwidth and self-energy of the 2D orbital (see Ref. \onlinecite{Louie1980periodic} for details). $\chi$ is not known \emph{a priori} in our case, but an agreement is found when using $\chi \approx 2.1$. From this, we find that the full width at half maximum (FWHM) is $\approx \SI{0.25}{\per\angstrom}$.

Building on the work by Mazzola \emph{et al.} \cite{Mazzola2014determining}, we use the intensity distribution in $\textbf{k}_\perp$ to estimate the real space width of the $\delta$-state. In our case, the electronic confinement in the Si:Sb $\delta$-layer is found to be \SI{1}{\nano\meter}. Although this is quite wide compared to the initial thickness of \ce{Sb} as deposited during sample preparation, it still represents a very narrow electronic confinement. Although the initial layer of Sb was about 1/3 of a monolayer, it is remarkable that the confinement remains this sharp in spite of the observed dopant segregation. \mb{The proximity of the surface could play a role in maintaining this level of confinement.} 
Importantly, the width of the electronic state being on the order of a nanometre is a clear sign that it indeed resides beneath the surface, and that the observed dopant segregation effect just manifests as an asymmetry and broadening of the dopant profile. This also means that $\delta$-layers made in this way are quite robust with respect to the particular shape of the confinement potential. 

The robustness of the Sb $\delta$-layer with the relatively simple, low-temperature fabrication process makes it quite promising with regards to potential applications in silicon-based electronics utilising reduced dimensions. Keeping post-growth annealing temperatures below \SI{450}{\celsius} means that Sb $\delta$-layer fabrication is compatible with established lower-temperature CMOS processing routines.
Moreover, Sb deposition by simple sublimation means that patterned $\delta$-layers could easily be made by utilising simple masking or straightforward lithography patterning techniques \cite{Rost2021simplified}, instead of having to rely on the extra surface treatment steps required in gas-phase deposition such as scanning tunneling microscopy (STM) \cite{Schofield2003atomically} or e-beam lithography \cite{Cooil2017insitu}. By further optimising growth parameters to facilitate fine tuning of the band structure, Sb $\delta$-layers could be made into indispensable building blocks in nanoscale electronics.

\section{Conclusion}
We have studied the occupied band structure of a \ce{Si}:\ce{Sb} $\delta$-layer by ARPES and DFT. We found that a very strong confinement potential exists such that the so-called $\Gamma$ state (the quantum state derived from the confinement of the bulk conduction band minima) is partially occupied and can be observed. This is in spite of the dopant layer evidently being partially segregated into the Si epilayer. The observed $\Gamma$ state has clear similarities with the extensively studied Si:P $\delta$-layer states, where the band is near-parabolic just below the Fermi level, and shows a slight difference in dispersion for the $\left[ 100 \right]$ and $\left[ 110 \right]$ directions. From photoemission studies, we found the confinement potential to be in the order of \SI{0.6}{\eV} (similar to Si:P $\delta$-layers). By probing the $\textbf{k}_\perp$-dependence of the photoemission intensity, we also found that the FWHM of the electronic state extended $\sim \SI{1}{\nano\metre}$. This is wide compared to the physical distribution of dopants even after segregation towards the surface, but still represents a relatively narrow confinement (c.f. 0.30 to \SI{0.68}{\nano\metre} for Si:P $\delta$-layers). Our observations demonstrate the robustness of the $\delta$-layer state against the particular thickness and symmetry of the dopant profile. These points serve to renew the candidacy of \ce{Sb} as a viable alternative to \ce{P} in $\delta$-doped semiconductors, while also being simpler and safer to handle during the fabrication process. Furthermore, Sb doping opens possibilities for increasing SOC and extending the quantum fabrication toolbox.  It is important to notice that similar challenges associated with dopant density and segregation were originally encountered with Si:P $\delta$-layer growth -- but these challenges were overcome by systematic optimisation of growth parameters (such as the substrate temperature). The obvious next step for the Si:Sb $\delta$-layer platform would be to perform a similar systematic optimisation and thus to improve the quality of Si:Sb $\delta$-layers and obtain more detailed information about the band structure.

\section{Acknowledgements}
This work was partly supported by the Research Council of Norway, project numbers 324183, 315330, 335022 and 262633. We acknowledge funding via the ‘Sustainable Development Initiative’ at UiO, as well as funding from Danscatt (7129-00011B). We also acknowledge MAX IV Laboratory for time on Beamline FlexPES under Proposal 20221393. Research conducted at MAX IV, a Swedish national user facility, is supported by the Swedish Research council under contract 2018-07152, the Swedish Governmental Agency for Innovation Systems under contract 2018-04969, and Formas under contract 2019-02496. 

This work was performed, in part, at the Center for Integrated Nanotechnologies, an Office of Science User Facility operated for the U.S. Department of Energy (DOE) Office of Science. Sandia National Laboratories is a multi-mission laboratory managed and operated by National Technology \& Engineering Solutions of Sandia, LLC (NTESS), a wholly owned subsidiary of Honeywell International Inc., for the U.S. Department of Energy’s National Nuclear Security Administration (DOE/NNSA) under contract DE-NA0003525. This written work is authored by an employee of NTESS. The employee, not NTESS, owns the right, title and interest in and to the written work and is responsible for its contents. Any subjective views or opinions that might be expressed in the written work do not necessarily represent the views of the U.S. Government. The publisher acknowledges that the U.S. Government retains a non-exclusive, paid-up, irrevocable, world-wide license to publish or reproduce the published form of this written work or allow others to do so, for U.S. Government purposes. The DOE will provide public access to results of federally sponsored research in accordance with the DOE Public Access Plan.

\section{Conflict of Interest}
The authors declare no competing interests. 

\section{Data Availability}
The datasets used in this study are available from the corresponding author upon reasonable request.

\section{Author Contributions}
F.S.S., S.P.C., J.J.F., H.I.R., A.C.Å., A.J.S., M.P.S., J.H., J.B., V.B., A.B.P., Z.L., M.B., J.A.M. and J.W.W. contributed to the data acquisition, and Q.T.C. performed DFT calculation. F.S.S. analysed the data with help from S.P.C. F.S.S. wrote the manuscript with input from J.W.W. and all the authors. The project was conceived and led by J.W.W.

\bibliography{Delta-layers}

\end{document}